\documentclass[preprints,review,accept,moreauthors,pdftex]{mdpi}

\firstpage{1} 
\pubvolume{xx}
\issuenum{1}
\articlenumber{5}
\pubyear{2019}
\copyrightyear{2019}
\history{\textit{Galaxies} 2019, 7(4), 92; \url{https://doi.org/10.3390/galaxies7040092}}

\pdfoutput=1

\usepackage{subcaption}
\usepackage{aascite}  
\usepackage{figs}     

\Title{Red Supergiants, Yellow Hypergiants, and Post-RSG Evolution}

\Author{Michael S.\ Gordon $^{1,\dagger,*}$\orcidA{} and Roberta M.\ Humphreys $^{2}$\orcidB{}}

\AuthorNames{Michael S. Gordon and Roberta M. Humphreys}

\address{%
$^{1}$ \quad SOFIA Science Center, NASA Ames Research Center; mgordon@sofia.usra.edu\\
$^{2}$ \quad Minnesota Institute for Astrophysics, School of Physics and Astronomy, 116 Church St.\ SE, University of Minnesota, Minneapolis, MN 55455, USA; roberta@umn.edu}

\corres{Correspondence: mgordon@sofia.usra.edu}

\firstnote{Current address: SOFIA Science Center, NASA Ames Research Center, Moffett Field, CA 940353} 

\abstract{How massive stars end their lives remains an open question in the field of star evolution.  While the majority of stars above $\gtrsim$9~$\mathrm{M}_\odot$ will become red supergiants (RSGs), the terminal state of these massive stars can be heavily influenced by their mass-loss histories.  Periods of enhanced circumstellar wind activity can drive stars off the RSG branch of the HR Diagram. This phase, known as post-RSG evolution, may well be tied to high mass-loss events or eruptions as seen in the Luminous Blue Variables and other massive stars. This article highlights some of the recent observational and modeling studies that seek to characterize this unique class of stars, the post-RSGs, and link them to other massive objects on the HR Diagram such as LBVs, Yellow Hypergiants, and dusty RSGs.}


\begin{document}



\section{Introduction}\label{sec:intro}
The standard model of massive star evolution follows a rapid progression from the main-sequence, through a blue supergiant (BSG) phase, to the red supergiant (RSG) branch, to terminal supernova (SN) explosion.  However, surveys of the brightest supergiants revealed an empirical upper luminosity limit to stars on the Hertzprung-Russell (HR) diagram \citep{humphreys1979}.  This limit suggests that stars above some initial zero-age main-sequence (ZAMS) mass ($\approx30-40\,\mathrm{M}_\odot$) do not evolve to the RSG branch on the HR Diagram, and therefore follow an alternative evolutionary pathway.  Since the 1980s, both observational and modeling studies have attempted to describe and constrain the stellar populations and instabilities at the upper luminosity boundary, as well as explore the local environments that influence these massive stars during both their main- and post-main-sequence lives.

It has long been established that massive stars at any stage of evolution provide a favorable environment for enhanced mass-loss in their stellar winds due to low surface gravity ($g$) in their outer atmospheres.  The outer circumstellar (CS) material is only tenuously gravitationally bound to the star itself.  Indeed, mass-loss rates for RSGs range from $10^{-6}\,\mathrm{M}_\odot\,\mathrm{yr}^{-1}$ \citep{gehrz1971,mauron2011} to as high as $10^{-4}\,\mathrm{M}_\odot\,\mathrm{yr}^{-1}$ in extreme supergiant stars like VY~CMa \citep{humphreys2013}---mass-loss rates that represent a significant fraction of a star's initial mass being shed during its post-main-sequence lifetime.  The evolution  and terminal state of a massive star is ultimately governed not just by its ZAMS mass but also by these drastic changes in total stellar mass and outer envelope conditions.  We refer to these changes in stellar mass through ejection of CS material as the ``mass-loss history.'' In this chapter, we summarize some of the literature on the mass-loss histories of evolved supergiant stars and the evidence for post-red supergiant evolution both in observational studies of the circumstellar ejecta and in evolutionary  models that predict the effect of various mass-loss mechanisms on massive star evolution.

\section{Context -- The Red Supergiant Problem}\label{sec:context}
Further context for much of the observational exploration of the last decade comes from a recently-identified ``red supergiant problem'' \citep{eldridge2008,smartt2009a,smartt2009b}.  A survey of Type II-P supernova progenitors using optical and near-IR pre-explosion archival images revealed an upper limit of only $16-17\,\mathrm{M}_\odot$ for the initial stellar masses of their likely red supergiant (RSG) progenitors  \citep{smartt2009a}.  Type II-P SN remnants are a useful laboratory for population statistics of RSGs since they represent the most abundant class of CCSNe ($\sim$70\% of hydrogen-rich SNe; see \citep{li2011} for a discussion of SN rate estimates).


From the notable lack of II-P SN progenitors above $\sim$17~$\mathrm{M}_\odot$ (Figure~\ref{fig:Smartt}), \citet{smartt2009a} suggested two possible scenarios:
\begin{enumerate}[leftmargin=*,labelsep=4.9mm]
\item	\label{enum:one}Systematic underestimation of progenitor mass due to improper extinction correction.
\item	\label{enum:two}Red supergiants greater than  $17\,\mathrm{M}_\odot$ have another terminal state besides II-P CCSNe.
\end{enumerate}
We explore these two scenarios below.

\figSmartt

Though no Type II-P SN progenitors appear to exist much above $\sim$16~$\mathrm{M}_\odot$, red supergiants have certainly been observed with masses far greater than this. For examples, see the recent HR Diagrams for massive evolved stars in the Galaxy \citep{levesque2005}, the Magellanic Clouds \citep{levesque2006}, and in M31 and M33 \citep{massey2009,drout2012,gordon2016}
A comparison with the evolutionary models \citep{meynet2003}, illustrates that many stars are present on the RSG branch above 20~$\mathrm{M}_\odot$ in Local Group galaxies.


One potential caveat for these RSG surveys is that the derived masses require accurate measurement of bolometric luminosity.  As \citet{smartt2009a} suggest in scenario~\ref{enum:one}.\ above, underestimating RSG masses could be a potential solution to the RSG problem.  Extra intrinsic extinction due to dust close to the RSG progenitor would yield lower luminosities and their estimated masses \citep{smith2011,walmswell2012}.  
Indeed, mass-loss rates and dust ejecta masses scale with luminosity \citep{van-loon2005,mauron2011,gordon2016}. Additionally, mid-IR interferometry around evolved stars has revealed dust close enough in to the central source \citep{danchi1994} that the dust grains could potentially be destroyed by the star's SN explosion.



\citet{walmswell2012} examined the effect of dust on derived RSG masses by applying the Cambridge STARS code \citep{eggleton1971,stancliffe2009} combined with various mass-loss schema \citep{de-jager1988,vink2001,eldridge2004} to simulate SEDs throughout a massive star's evolution.  They model a series of dust shells and estimate a simulated extinction to measure an average $A_V$ of around 1~mag from the circumstellar ejecta.  The resulting model SEDs do indeed yield lower deduced stellar masses than what would have been observed with proper extinction-correction for the model dust shells, with an underestimate of as much as 5~$\mathrm{M}_\odot$ for supergiants in the $\sim$20$-25\,\mathrm{M}_\odot$ range.
Still, the authors note that even a change in several solar masses worth of dust material does not solve the red supergiant problem.




\citet{beasor2016} explored this missing mass problem  combining mid-IR WISE and Spitzer/IRAC photometry with circumstellar dust shell models from DUSTY \citep{ivezic1999}. The authors apply their model analysis to a co-eval population of RSG cluster stars (NGC 2100), which allows for studying stars with similar initial conditions---mass, metallicity, local environment, etc.  If the cluster stars all have roughly the same initial masses (within a few tenths of a solar mass), then the evolutionary pathway should be the same, and any differences in luminosity should be due to the slightly more massive stars evolving faster on the HR Diagram. This allowed the authors to use luminosity as a proxy for evolution.  Based on their models and estimated mass-loss rates, they find an increase in mass-loss rates along the RSG branch as high as a factor of 40 over the post-main-sequence lifetime of the star, which appears to be consistent with the \citet{de-jager1988} mass-loss prescription. If the increased mass loss is translated into an intrinsic exctinction, they argue that the increased reddening may substantially increase derived masses of Type II-P SN RSG progenitors. As an example, the authors show that similar dust extinction conditions on SNe 1999gi, 2001du, and 2012ec could revise initial mass estimates by as many as 10~$\mathrm{M}_\odot$.

\citet{kilpatrick2018}, however, argue that while circumstellar dust can  alter the observed SEDs of supergiant progentors, several studies of the circumstellar environments around SN progenitors suggest that there cannot be enough material around at least some SN Type~II progenitor systems to hide an underlying high-mass RSG. 
They note that the total dust mass in progenitor systems is independently constrained by radio and X-ray observations.  For example, X-ray light curves of CCSNe have been used to estimate stellar wind parameters and the density structure of the CS medium \citep{chevalier2003,dwarkadas2012}. Many of these studies, though, have broad wavelength coverage of the SN progenitor SED.  It is possible that for some RSG progenitors with less constrained SEDs and sparse pre-SN imaging/photometry, the ``missing mass'' scenario from CS dust may indeed be biasing RSG mass statistics.  However, as IR photometry exists for many SN RSG progenitors, this argument is only a partial solution to the red supergiant problem.

As for scenario~\ref{enum:two}.\ above from \citet{smartt2009a}---that high-mass red supergiant progenitors simply do not exist---there are two possible explanations: first, that higher mass RSGs collapse directly to black holes; and second, that stellar evolution to the warmer, blue side of the HR Diagram produces stellar end products other than Type II-P SNe.  The subject of black hole formation, either through direct collapse or fall back, merits a longer discussion that is beyond the scope of this work.  For a review of some of the work surrounding black hole formation in massive stars, see the annual review by \citet{smartt2015}.

One realm of exploration in the literature is the idea of failed supernovae, or ``unnova'': stars that collapse to black holes with little or no energy released \citep[e.g.,][]{kochanek2008,ugliano2012,pejcha2015,gerke2015}). Such events may have no significant transient, and thus be almost impossible to observe \citep{woosley2012}. However, models by \citet{lovegrove2013} and \citet{piro2013} find that RSGs in the $15-25\,\mathrm{M}_\odot$ range can lose so much energy in neutrinos during collapse that the resulting shock in the stellar envelope is expected to create an optical signature.  This can be as bright as $L_{\mathrm{bol}}\sim10^6-10^7\,L_\odot$, though perhaps lasting for only a few days \citep{piro2013}.


These results suggest that an optical transient of this type from a failed SN would have only a small observable window, thereby decreasing the likelihood of detection.  Nonetheless, surveys like that on the LBT \citep{kochanek2008,gerke2015,adams2017} have potentially found one such source, N6946-BH1, which brightened to $\gtrsim10^6\,L_\odot$ in March 2009 before fading below its pre-outburst luminosity \citep{gerke2015}. SED modeling constrained the mass of the RSG progenitor to $\sim$25 $\mathrm{M}_\odot$ \citep{adams2017}, above the apparent \citet{smartt2009a} SN Type II-P progenitor limit.  Despite a decade of monitoring, however, objects like this remain exceedingly rare.  While failed SNe may indeed represent some high-mass RSG population that is as of yet undiscovered, for the moment this does not seem to solve missing high-mass SN progenitors.

In this chapter, we focus on another population of transient objects---the yellow supergiants (YSGs) and hypergiants, and evidence for post-red supergiant evolution.

\section{The Milky Way Hypergiants and Post-RSG Evolution}

Many years before attention was drawn to the red supergiant problem, a small group of high luminosity evolved supergiants was recognized with a range of intermediate to cool temperatures, high mass loss, and unstable atmospheres \citep[see several papers in][]{de-jager1992}.  These stars, now referred to as yellow or red hypergiants, defined the empirical upper luminosity boundary in the HR Diagram for evolved massive stars \citep{humphreys1979}.  The well-studied members of this elite group are Galactic members with a few examples in the Magellanic Clouds and M31 and M33 (\S\ref{sec:ysgs}). The visibly bright Galactic stars all exhibit spectroscopic and photometric variability, high mass loss, and several show dusty ejecta.  The evolutionary state of the warm or yellow hypergiants was not obvious; they could be evolving toward the red supergiant region or on a blue-loop back to warmer temperatures.  The instability and brief high mass-loss events exhibited by $\rho$~Cas, for example \citep{lobel2003}, during which it developed TiO bands, and the increasing evidence for episodic high mass-loss events especially visible in the ejecta of IRC~+10420 (\S\ref{sec:irc}) favored a post-RSG evolved state for these stars.

Other warm or yellow hypergiants include the Galactic stars HR~8752 and HR~5171A \citep{nieuwenhuijzen2012,lobel2015,chesneau2014}. These hypergiants are visually bright and relatively nearby, which has made them important laboratories for study of late-stage evolution. Interestingly, \citet{de-jager1998} suggests that all of the yellow hypergiants are post-red supergiants. During blueward evolution, their atmospheres contract, the atmospheric opacity increases, and their rotation increases.  Having shed a sizable fraction of their mass on the RSG branch, these stars are now closer to the Eddington Limit for their ZAMS mass. The stars thus enter a temperature range (6000--9000~K) of increased dynamical instability, that de~Jager called the ``yellow void,'' where high mass-loss episodes occur.  Figure~\ref{fig:void} is a schematic HR Diagram showing the positions of some of the better-studied Galactic yellow hypergiants plus Var~A in M33 with respect to the critical temperature region.

\figVoid

\section{IRC~+10420 and Var~A in M33 -- Clues to Post-RSG Evolution}\label{sec:irc}
The luminosities and apparent temperatures of the two evolved yellow supergiants IRC~+10420 and Var~A place them at the upper luminosity boundary for evolved stars in the HR Diagram. Both stars exhibit a history of photometric and spectroscopic variability with high mass loss episodes and dusty circumstellar ejecta making them excellent candidates for post-red supergiant evolution.

At its initial discovery, IRC~+10420 (V1302~Aql) was quickly recognized as remarkable with its very large infrared excess and late F-type high luminosity spectrum \citep{humphreys1973}. It was soon identified as a powerful maser source and is one of the warmest known OH/IR stars \citep{giguere1976}.

IRC~+10420 is a Galactic star and because of its relative proximity, its circumstellar ejecta is easily resolved by HST imaging \citep{humphreys1997} which revealed a complex environment. The color image in Figure~\ref{fig:IRC} shows the spatial extent of the ejecta, more that 5~arcsec across. Numerous features are visible within two arcsec of the embedded star including condensations arrayed in jet-like structures, rays, and an intriguing group of small, nearly spherical shells  or arcs apparently at the ends of some of the jet-like features.  One or more distant reflection shells at 5 to 6~arcsec from the star are visible in the longer exposure images.

\figIRC

While its actual distance is somewhat uncertain, numerous arguments \citep[e.g.,][]{oudmaijer1996} clearly demonstrated that it was not a post-AGB star and therefore above the AGB-limit at $\textrm{M}_{Bol}$ $\approx -7$~mag.  The reddening of its optical spectral energy distribution, infrared polarization, and its radial velocity \citep{jones1993} suggested a distance of 4--6 kpc and a luminosity of $\approx -9.6 \pm 0.5$~mag (at 5~kpc), which places IRC~+10420 at the upper luminosity boundary in the HRD for evolved stars. Jones et al. \cite{jones1993}  therefore proposed that IRC~+10420 may be evolving from a red supergiant across the HR Diagram to warmer temperatures, and in a phase of its evolution analogous to the post-AGB lower mass giants evolving to the planetary nebula phase, but at much higher luminosities.

The early photographic image-tube spectra \citep{humphreys1973} showed late F-type absorption features, however, 23 years later, Oudmaijer et al. \cite{oudmaijer1998} identified H lines and other absorption features typical of a warmer A-type supergiant implying a significant change in its apparent temperature. HST/STIS spectra a few years later were consistent with the higher temperature \citep{humphreys2002}. The spectrum is dominated by a strong H$\alpha$ stellar wind split emission line (Figure~\ref{fig:IRC}, right) due either to a bi-polar outflow or an equatorial disk. Strong Ca~II triplet emission lines, also with a split profiles, and the  [Ca~II] doublet in  emission formed in the extended low density ejecta plus numerous Fe~II emission lines typical of a stellar wind  are present. \citet{humphreys2002} demonstrated that the wind was optically thick. Thus, observed variations in the apparent spectral type and the inferred temperature are changes in the wind and not to changes in the interior, i.e.\ evolution, of the star on such a short timescale. Subsequent spectroscopic monitoring by \citet{klochkova2016} and by our group do not show any further increase in its apparent temperature suggesting that the blueward motion of IRC~+10420 on the HR Diagram  has slowed.

The morphology of IRC~+10420's circumstellar ejecta had always been elusive, with suggestions of a bipolar outflow or a circumstellar disk with different orientations ranging from edge-on to an inclined disk at different angles. To investigate its three-dimensional morphology, \citet{tiffany2010} combined the transverse velocities for several knots and condensations in the inner ejecta, measured from second-epoch HST imaging, with their Doppler velocities.  The resulting total space motions and direction of the outflows showed that these knots were ejected at different times and in different directions over the last $\approx400$~years, a relatively recent period of asymmetric mass loss. Interestingly, they are all moving within a few degrees of the plane of the sky. Thus we are viewing IRC~+10420 nearly pole-on and are looking nearly directly down onto its equatorial plane. This orientation is confirmed by both the highly polarized 2.2~$\mu$m emission around the star, which places the scattering dust in the plane of the sky \citep{shenoy2015}, and high resolution near-infrared interferometry \citep{de-wit2008}. The more distant reflection shells were ejected about 3000 years ago, suggesting more than one epoch of high mass loss.

To explore IRC+10420's mass loss history, \citet{shenoy2016} used far-infrared imaging from SOFIA/FORCAST at 11--37~$\mu$m to probe the extended cold dust plus high-resolution adaptive optics imaging at 8--12~$\mu$m. They found evidence for two distinct periods of high mass loss, an earlier episode from 6000 to about 2000 years ago with a high rate of 2 $\times 10^{-3}\,\mathrm{M}_{\odot}\,\mathrm{yr}^{-1}$, followed   by an order of magitude decrease  with a current rate of $\approx$ 10$^{-3}\,\mathrm{M}_{\odot}\,\mathrm{yr}^{-1}$, consistent with other recent measurements. This change is additional evidence for IRC+10420's evolution from the red supergiant stage and its transition to a warmer state.

Var~A in M33 is significant since it has actually been observed to transition to a red supergiant and back to its presumably normal state as a high luminosity F-type supergiant within the last century. This color and spectral change, however, was not due to interior evolution, but to a high mass-loss episode that produced a dense, cooler wind. Var~A provides additional evidence for the highly unstable state of evolved stars near the upper limit in the HR Diagram.  This supergiant has the important advantage that its distance and therefore its intrinisic luminosity, M$_{Bol}\approx -9.5$, are known. In M33, however, it is too distant for direct imaging of its ejecta.

Var~A is one of the original Hubble-Sandage (H-S) variables \citep{hubble1953}. However, unlike the other H-S variables that have been subsequently identified as evolved hot stars with episodes of high mass loss---the LBVs---Var A's quiescent state is a high luminosity yellow or intermediate temperature supergiant.  Its historic light curve \citep[Figure~6 in][]{hubble1953} is remarkable. At maximum light it was one of the visually-brightest stars in M33, but then in 1951 its luminosity rapidly declined by 3.5~mag, becoming faint and red after what had been a slow increase in brightness during the previous 50 years.   Spectra from 1985 and 1986 revealed an M-type supergiant with prominent TiO bands \citep{humphreys1987}. Its spectral energy distribution not only showed the shift to cooler temperatures but a large mid-infrared excess due to extensive circumstellar dust, and Var~A was as luminous at 10~$\mu$m as at its visual maximum. The star had experienced a high mass-loss event that had produced an optically thick, cooler wind---a ``false'' or ``pseudo'' photosphere---that resembled a red supergiant.

Subsequent spectra, not observed until 2003--2004, revealed that its ``eruption,'' which had begun $\sim$1951, had indeed ended, having lasted $\approx 45$~years \citep{humphreys2006}. The spectrum showed that the star or its dense wind was now in a much warmer state with absorption lines consistent with an F-type supergiant and emission lines of Ca~II, [Ca~II] and K~I, similar to IRC~+10420, in addition to strong H emission formed in it surrounding low-density gas. The optical photometry shows the transition to bluer, warmer colors, but Var~A remained visually faint and was still obscured by circumstellar dust.  The spectra from 1985 and 2004 are shown in Figure~\ref{fig:VarA} and its light curve and SED in Figure~\ref{fig:VarAlc}. Its 10~$\mu$m flux shows an unexpected decline, which implies an unexpected decrease in the star's total luminosity. The most likely explanation is that the radiation is escaping in some direction other than our line-of-sight. This possibility is supported by recent spectra of small clumps and knots in the inner ejecta of the red hypergiant VY~CMa \citep[][see \S\ref{sec:massloss} below]{humphreys2019}, which require a clear line of sight to the star, and therefore imply large, low density regions even holes in the circumstellar material which may also be the case for Var A. 

\figVarA
\figVarLC

Thus, Var~A and IRC~+10420 are not only probable post-red supergiants, but their shared characteristics of photometric and spectroscopic variability, surface instability and stellar winds, high mass loss, and a history of enhanced mass-loss episodes are clues to understanding the evolution of stars near the upper luminosity boundary and their transit across the HRD from red to blue.

\FloatBarrier

\section{YSGs and Post-RSG Candidates in M31 and M33}\label{sec:ysgs} 

Other than Var~A, IRC~+10420, and the Galactic hypergiant candidates, what fraction of known evolved supergiants may be in a post-RSG state?  These statistics, as well as the physical characteristics of candidate post-RSGs and their locations on the HR Diagram, are crucial to our understanding the final stages of the majority of massive stars.

Due in part to their position on the HR Diagram, few post-RSGs are known. They occupy a relatively brief, transient state between the blue and red supergiants and may either be evolving from the main sequence to cooler temperatures, or back to warmer temperatures from the RSG stage. In the Galaxy, the warm or yellow hypergiants, close to the upper luminosity boundary in the HRD with high mass-loss rates, enhanced abundances, and dusty CS environments, are excellent candidates for post-RSG evolution. These stars contrast with the intermediate-type yellow supergiants which have normal spectra and long-wavelength SEDs---that is, no evidence for circumstellar dust or mass loss in their spectra.  Considering how few objects of this type are known locally, many studies have pursued observations of supergiants outside of the Galaxy.

As part of a larger program on the luminous and variable emission-line stars in M31 and M33, \citet{humphreys2013} recognized a few high luminosity, A- to F-type stars in each galaxy with spectroscopic evidence for high mass loss, and extensive gaseous and dusty circumstellar ejecta revealed in their spectra and SEDs; characteristics shared with the warm hypergiants IRC~+10420 and the peculiar Var~A also in M33. They  demonstrated that these stars were indeed evolved, intermediate temperature supergiants with strong winds and mass loss, and like IRC~+10420 and Var~A, they were candidates for post-RSG evolution. Based on their luminosities, their initial masses would be greater than 20~$\mathrm{M}_\odot$ or more. One possible exception was B324, one of the visually brightest stars in M33. Its SED showed strong free-free emission in the near-infrared but lacked the cooler dust expected in a post-RSG star.  B324, just at the upper luminosity boundary, with high mass loss, could be approaching the limit to its redward evolution and therfore a candidate for future high mass episodes.  \citet{humphreys2013}   had identified a few candidates, but was not a comprehensive survey for post-red supergiants. 

\citet{gordon2016} conducted a survey of the yellow and red supergiants to search for post-RSG candidates. The targets were primarily selected from the published surveys of M31 and M33 for yellow and red supergiants \citep{drout2009,drout2012,massey2009} chosen from  the Local Group Galaxies Survey (LGGS; \citep{massey2006}).  Post-RSG candidates were identified based on spectroscopic evidence for mass loss and the presence of circumstellar dust in their SEDs. In that work, \citet{gordon2016} spectroscopically confirmed 75 YSGs in M31, 30 of which (40\%) are likely in a post-RSG state based on spectroscopic and photometric markers for dusty wind.  For M33, 27 of the observed 86 YSGs (31\%) were determined to be post-RSG candidates.  Further discussion of this work and its methodologies is included below.  We note that a similar survey was conducted by \citet{kourniotis2017}, which flagged yellow super and hypergiant candidates based on photometric criteria for follow-up spectroscopy.

The greatest challenge in photometric surveys of supergiants is distinguishing extragalactic sources from foreground disk dwarfs as well as halo giants in the Milky Way. \citet{humphreys1980}  highlighted the magnitude of this issue in a survey of the brightest blue and red supergiants in M33. There is significant contamination of foreground K and M dwarfs in the red supergiant region of the M33 color-magnitude diagram (CMD), which presents some observational challenges.  Since there is little star formation in the Milky Way halo, there is essentially no foreground contamination in the ``blue plume'' of the CMD. \citet{massey2007} applied the \citet{bahcall1980} model to estimate that almost 80\% of red stars ($1.2 < B - V < 1.8$) fainter than $V\sim16$ seen toward M31 will be foreground stars. The central portion of the CMD, representing the yellow supergiant population, is similarly affected by foreground contamination.  \citet{drout2009} and later \citet{massey2016} apply the Besan\c{c}on model \citep{robin2003} of the Milky Way (two disks + halo) to illustrate that over 70\% of bright stars redward of the blue plume ($B - V > 0.4$) could be foreground contamination. 

Massey et al. \cite{massey1998,massey2009} and \citet{drout2012} demonstrated that color criteria could be used as an effective metric for distinguishing foreground contaminants in the RSG surveys in M31 and M33, but few such two-color discriminants have been used for YSGs, except for Bonanos et al. \cite{bonanos2009,bonanos2010}, who defined color ranges for a variety of massive star types in the Magellanic Clouds using 2MASS and Spitzer/IRAC photometry. In general, however, spectra are needed to determine both extragalactic membership and evolutionary state.

\subsection{Spectral Types and Luminosity Classification}

\citet{drout2009,drout2012} use radial velocities from spectral-line features to generate a catalog of extragalactic YSG candidates, whereas both \citet{gordon2016} and \citet{massey2016} classified the stars based on the spectral type and luminosity criteria in their absorption-line spectra. For example, the blends of Ti~II and Fe~II at $\lambda\lambda4172$-8 and $\lambda\lambda4395$-4400 are valuable luminosity criteria in the blue when compared against Fe~I lines, which show little luminosity sensitivity such as $\lambda4046$ and $\lambda4271$.  The O~I~$\lambda7774$ triplet in the red spectra---also used in \citet{drout2012} as part of their classification scheme---is also a particularly strong luminosity indicator in A- to F-type supergiants. 

Using these and several other classifiers, \citet{gordon2016} confirmed extragalactic membership of $\sim$150 yellow supergiants in M31 and M33. Thirty, or $\sim$20\%, of the observed YSGs in each galaxy showed evidence for stellar winds in their ejecta and enhanced mass-loss, not shared with the other YSGs, and therefore possible post-RSG evolution.  The notable spectral features include P~Cygni profiles in hydrogen emission, broad wings in H$\alpha$ or H$\beta$ emission indicative of Thomson scattering, and [Ca~II]/Ca~II triplet emission from circumstellar gas. 
If mass-loss markers in the YSG SEDs are included (discussed below), the fraction of YSGs likely in a post-RSG state increases to $\sim$40\% of the observed sources.

\subsection{Photometric Evidence of Mass Loss}
\citet{gordon2016} also examined the SEDs of the YSG and RSG  populations in M31 and M33 to identify what fraction of the evolved supergiants have circumstellar dust and are in a mass-losing state.  The RSGs currently experiencing episodes of high mass loss may eventually evolve to become post-RSG warm supergiants, LBVs, or WR stars.

The defining signature of mass loss in RSGs is the presence of circumstellar dust, usually revealed as excess radiation in their IR SEDs from the silicate emission features at 9.8~$\mu$m and 18~$\mu$m, corresponding to the Si--O vibrational \citep{woolf1969} and O--Si--O bending modes \citep{treffers1974}, respectively. The strength of the silicate emission feature is (to first order) correlated with the luminosity and apparent temperature as revealed by the spectral type; i.e., the higher the luminosity and cooler the star, the stronger the silicate emission and the larger the IR excess.

In the YSGs, the presence of excess radiation due to circumstellar warm dust and/or free-free emission in the near and mid-infrared wavelengths is evidence for mass loss. This additional radiation is apparent in their SEDs if the flux in the near-IR bands exceeds the expected Rayleigh-Jeans tail of the stellar component. For example, an infrared excess in the $1-2\,\mu$m 2MASS bands is a well-known characteristic  of free-free emission in stellar winds, while the 3.6 to 8~$\mu$m Spitzer/IRAC data provides evidence for warm CS dust. Free-free emission is generally identified as constant F$_{\nu}$ in the near-infrared, often extending out to 5~$\mu$m. Examples are shown in Figure~\ref{fig:yhgsed} for two warm hypergiants in M31. Beyond being useful for identifying mass loss, this IR excess in the stellar SED is crucial for accurately calculating the bolometric luminosity.  The CS dust will re-radiate the central star's optical flux into the infrared, and this processed radiation can contribute significantly to the total bolometric luminosity of the star + ejecta system.  There are various methods for fitting models to stellar SEDs to account for this, and an example can be found in \citet{kourniotis2017}, who fit an ATLAS9 stellar atmosphere model \citep{kurucz1979,howarth2011} for the stellar component and up to three distinct blackbodies for the warm and cool dust components of their YSG SEDs.
\figYHGsed

\citet{gordon2016} find $\sim$50--60\% of the observed RSGs in M31 and M33 show evidence for an IR excess in their near- to mid-IR SEDs. The IRAC 8~$\mu$m photometry is used in \citet{gordon2016} to provide an estimate of the total dust mass lost over a timescale of about a century and estimate that the RSGs in both galaxies tend to have dusty ejecta of the order of $10^{-3}-10^{-2}~\textrm{M}_\odot$, assuming a warm dust component of $\sim$350~K.  Consistent with the \citet{de-jager1988} prescription, mass loss correlates with luminosity along the RSG branch. If more than 50\% of RSGs are indeed experiencing sufficient mass loss to produce CS dusty ejecta, a large fraction of stars along the RSG branch may evolve back toward the blue to become the warm post-RSG stars before their terminal state as SNe or black holes.

We note that the target selection from \citet{gordon2016}  was derived from optical surveys.  Thus, it may be likely that our surveys of the most luminous stars in M31 and M33 do not necessarily include some supergiant populations that are heavily obscured. Since the most luminous warm and cool supergiant populations are  likely to have the highest mass-loss rates, it is probable that some will be  obscured in the optical by their own CS ejecta in the optical surveys. To complete the upper portion of the HRD would require a further search in the IR to find the brightest infrared sources. There are several IR surveys of M31 and M33 with Spitzer/IRAC \citep[e.g.\ ][]{mcquinn2007,mould2008,khan2015} that have specifically targeted the bright and/or variable stellar populations in the Local Group.  These surveys have already revealed many unique supergiant stars that were obscured in high-resolution optical surveys--for example, the discovery of optically-obscured $\eta$~Carinae analogs by \citet{khan2015b}.

\subsection{The Post-RSG Candidates, the HR Diagram, and Comparison with Evolutionary Models}
The HR Diagrams for the observed YSGs and RSGs in M31 and M33 \citet{gordon2016} are reproduced in Figure~\ref{fig:M33hrd}. For the YSGs with observed optical spectra, effective temperatures can be derived through comparison to intrinsic colors of the stars' identified spectral types.  However, for sources without observed spectra/spectral-type, several photometric temperature scales exist in the literature.  For example, \citet{massey2009} compare the $(V-K)$ colors of their M31 RSGs to MARCS atmosphere synthetic photometric colors, and \citet{drout2012} adopt the $(V-R)$ color transformations from LMC sources \citep{levesque2006} for their observed RSGs in M33.  We note that in the absence of spectral types, photometric temperature scales can be somewhat uncertain.
\figGordonHRD

In both M31 and M33, the post-RSG candidates---flagged in \citet{gordon2016} based on their spectroscopic and/or photometric mass-loss indicators---are preferentially more abundant at higher luminosities. Also shown in Figure~\ref{fig:M33hrd} are Geneva Group \citep{ekstrom2012,meynet2015} evolutionary tracks for different ZAMS mass models. The higher mass models ($\mathrm{M}\gtrsim20\,\mathrm{M}_\odot$) loop through the YSG region of the HRD, perhaps even in multiple passes, before terminating on the RSG branch.  These stars are those supergiants undergoing post-RSG evolution and are sometimes referred to in the literature as ``group 2 blue supergiants'' \citep[e.g.,][]{saio2013,georgy2014}.  We loosely define the YSG region as $\sim$4000 to $12\;000$~K, and this evolution across the HRD can occur over timescales of just a few Myr.

Comparison with the evolutionary tracks suggests that most of the progenitor main-sequence stars have masses $\gtrsim$20~M$_\odot$. Likewise, the dusty RSGs dominate the higher luminosities. This is not surprising considering results from \citet{mauron2011} (Figure~\ref{fig:mauron}) and others that $\dot{\mathrm{M}}$ and total mass lost in the RSGs correlates with luminosity.

HR Diagrams of massive stars in the Local Group like those in \citet{gordon2016} and others \citep[e.g.\ ][]{levesque2006,massey2009,massey2016,drout2012,meynet2015a,kourniotis2017,kourniotis2018} suggest that the mass-losing post-RSG candidates are more common at luminosities above $\sim$10$^5\,L_\odot$. Most appear to have initial masses of 20--40~M$_\odot$, and may be the evolutionary descendants of the more massive RSGs that do not explode as supernovae (i.e.\ the ``missing'' RSGs from \citet{smartt2009a}). The eventual fate of these stars may be either as ``less-luminous'' LBVs or WR stars before their terminal explosion.


\section{Mass-Loss in the Yellow and Red Supergiants}
For many YSG and RSG stars, the \textit{thermal} excess flux is fairly constant across the mid-infrared, which implies that the dust is emitting over a range of temperatures and distances from the central star.  With some assumptions on dust temperature, grain size distributions, silicate grain chemistry, and gas-to-dust ratio, near- to mid-infrared photometry can be used directly to estimate the total mass of the CS ejecta  around each supergiant star.  With some additional measurements and/or assumptions on timescales---such as the stellar wind velocity \citep{humphreys2013,gordon2016}, or the dust condensation timescale---estimates on mass-loss rates can be extracted from the mid-infrared flux alone.  For example, \citet{mauron2011} apply the \citet{de-jager1988} mass-loss prescription to Galactic RSGs to estimate an average mass-loss rate of $\sim$10$^{-6}\,\mathrm{M}_\odot\,\mathrm{yr}^{-1}$ from IRAS 60~$\mu$m flux.  Figure~\ref{fig:mauron} from \citet{mauron2011} illustrates the \citet{de-jager1988} prediction of increasing mass-loss rate with increasing luminosity for a handful of Galactic RSGs.  Similar figures exist in \citet{meynet2015a,gordon2016} and others for Galactic and extragalactic RSGs (see Figure~\ref{fig:gordon} below which illustrates a similar trend for \textit{total} ejecta mass lost).
\figMauron
\figGordon

\FloatBarrier


The DUSTY radiative transfer code \citep{ivezic1999} is now often used to derive mass-loss rates or total ejected mass.  DUSTY solves the radiative transfer equation for a spherically-symmetric dust distribution around a central source. Input parameters include the spectrum of the illuminating source, the optical properties and size distribution of the dust grains, the dust temperature at the inner boundary of the shell, and a functional form for the radial profile of the dust density throughout the shell.  The primary output is the resulting SED of the modeled system. This code has been recently been applied to different populations of RSGs and their ejecta to derive $\dot{\mathrm{M}}$-luminosity relations from the IR SED fitting \cite{beasor2016,beasor2018,shenoy2016,gordon2018}.

\citet{shenoy2016} and \citet{gordon2018} used DUSTY to generate radial profiles for a variety of model dust density profiles to test whether the mass-loss rates of the target RSGs are constant and smooth over time (e.g., $\rho_\mathrm{dust}\propto r^{-2}$), or if the circumstellar ejecta can be better modeled by one (or more) discrete, high-mass ejecta events. 
This methodology, however, requires high-resolution imaging both to trace the ejecta close to the central star and also to resolve the thermal emission above the PSF of the telescope/instrument used for the observations. These studies demonstrated that a spherically-symmetric shell model with constant mass loss over time does not adequately explain the morphology of the circumstellar ejecta in many yellow and red supergiants. In fact, variable mass loss over time is required to build up the multiple dust shells observed around several Galactic RSGs.

\subsection{Mass-loss Mechanisms and High Mass-Loss Events}\label{sec:massloss}
Both ground and space-based high-resolution imaging and interferometry of evolved massive stars are transforming our view of  mass loss and the mass-loss mechanism in evolved stars. The precise mass-loss mechanism for red supergiants is not fully understood.  The leading processes have included radiation pressure on grains, pulsation, and convection.  The discovery of large-scale surface asymmetries or hot spots on the surfaces of red supergiants (\citep{gilliland1996,tuthill1997,monnier2004} and more recently \citep{montarges2018,nance2018,lopez-ariste2018}), which vary on short timescales of months or years, supports the important role of convection and surface activity.

Pulsation and dust-driven winds have been successful at explaining mass loss in Miras and AGB stars, which are fundamental-mode pulsators.  However, less variable RSGs with extended, low-density atmospheres are quite different environments than their lower-mass counterparts. Pulsation may be important for the YSGs, which are at the upper-luminosity limit of the Cepheid instability strip. For example, the light and velocity curves for $\rho$~Cas \citep{lobel2003} support a pulsational instability as the origin of its three brief, high mass-loss episodes. Yet, as discussed in \S\ref{sec:irc}, the peculiar M33 Var~A's 45+ year high mass-loss episode \citep{humphreys2006}, during which it resembled an M supergiant, required some high mass-loss mechanism lasting decades.  Additionally, there exists significant dispersion in the measured mass-loss rates for stars of a given luminosity class.  For example, \citet{mauron2011} compiled mass-loss rates for LMC RSGs from several data sets \citep{reid1990,van-loon2005,groenewegen2009} to demonstrate that for stars around $10^5$~L$_\odot$, a rather wide range of mass-loss rates between $10^{-6}$ and $10^{-4}\,\mathrm{M}_\odot\,\mathrm{yr}^{-1}$ have been measured (see their Figure~5).

This dispersion may well be due to observational bias or different measurement techniques, or may indeed be a manifestation of whatever physical mass-loss mechanism is at play. One approach to mitigate systematics in this mass-loss rate dispersion is to study individual RSG cluster populations.  \citet{beasor2018} compared RSGs within NGC~7419 and $\chi$~Per, whose stars are of similar ages \citep[$\sim$14~Myr;][]{currie2010,marco2013}. With a focus on these coeval populations, the effects of age, metallicity, and environment on $\dot{\mathrm{M}}$ are removed, and they  find a tight correlation of mass-loss rate with luminosity.

Optical and near-IR imaging of the extreme OH/IR supergiant VY~CMa and the post-RSG IRC~+10420 (\S\ref{sec:irc}) have yielded surprising results about the circumstellar environments around massive stars.  VY~CMa has an extensive, highly structured nebula consisting of multiple knots and arcs ejected within the last 1000 years \citep{humphreys1997,humphreys2005,humphreys2007,smith2001,tiffany2010}. The numerous knots, arcs, and loops visible in scattered light from the dust in their ejecta are structurally and kinematically distinct from the surrounding diffuse ejecta (see, for example, \citealt{smith2001,humphreys2019}). These features were each ejected at different times over several hundred years, presumably by localized processes from different regions on the star. Estimates of the mass in some of the arcs and clumps in VY~CMa's ejecta from surface photomety in the HST images and from the near-IR imaging of the southwest clump feature \citep{shenoy2013,gordon2019}, yield minimum masses of 3--5$\times10^{-3}\,\mathrm{M}_\odot$ implying short term, high mass-loss events. These discrete ejecta events hint at a very different ejecta mechanism than the slow, spherical shell paradigm. The presence of magnetic fields from Zeeman splitting and polarization of the OH/water masers has been detected in the circumstellar ejecta of VY~CMa and other OH/IR supergiants such as VX~Sgr, NML~Cyg, and S~Per \citep{vlemmings2002,vlemmings2003,vlemmings2005}. These results suggest that enhanced surface convective activity \citep[e.g., in $\alpha$~Orionis;][]{kervella2016,nance2018,lopez-ariste2018} together with magnetic activity may be important for these high mass ejection events.

Recently, HST/STIS spectra revealed TiO and VO molecular emission discrete ejecta close to the central star in VY~CMa \cite{humphreys2019}.  These molecules, previously believed to form in low-density dusty CS shells, instead appear concentrated in small clumps and knots.  Coupled with extremely strong K~I emission \citep[4~L$_\odot$ in just two narrow doublet lines;][]{humphreys2005,humphreys2019}, the emission features imply a dust-free environment between the knots and the star. These localized sources of atomic and molecular emission imply major gaps or holes in the star's envelope or outflow structure perhaps formed by large-scale surface activity. 

Thus many of the luminous warm and cool hypergiants have extensive CS ejecta and evidence for high mass loss events. The yellow hypergiants and many of the yellow supergiants are candidates for post red supergiant evolution. IRC~+10420, Var~A, and the extreme red supergiant VY~CMa may be the special cases that provide the clues to understanding evolution near the top of the HR Diagram.  These stars represent  short-lived, unstable states that signal the last stages in RSG evolution and the brief post-RSG transition as the star returns to warmer temperatures.  This class of post-RSG stars with complex mass-loss histories may be the missing piece on the HR Diagram  and the solution to the red supergiant problem.

\FloatBarrier

\vspace{6pt} 

\abbreviations{The following abbreviations are used in this manuscript:\\

\noindent 
\begin{tabular}{@{}ll}
  2MASS & Two-Micron All Sky Survey\\
  BSG & Blue Supergiant\\
  CCSNe & Core-collapse supernovae\\
  CMD & Color-magnitude diagram\\
  CS & Circumstellar [ejecta]\\
  HRD & Hertzsprung-Russell diagram\\
  HST & Hubble Space Telescope\\
  IR & Infrared\\
  IRAC & Infrared Array Camera (Spitzer)\\
  IRAS & Infrared Astronomical Satellite\\
  LBV & Luminous Blue Variable\\
  LGGS & Local Group Galaxy Survey\\
  LMC & Large Magellanic Cloud\\
  MIRAC & Mid-Infrared Array Camera (MMT)\\
  PSF & Point spread function\\
  RSG & Red Supergiant\\
  SED & Spectral energy distribution\\
  SN[e] & Supernova[e]\\
  STIS & Space Telescope Imaging Spectrograph (HST)\\
  WISE & Wide-field Infrared Survey Explorer\\
  WR & Wolf-Rayet\\
  YSG & Yellow Supergiant\\
  ZAMS & Zero-age main-sequence
\end{tabular}}


\reftitle{References}





\end{document}